\documentclass{aa}
\usepackage{epsfig}
\usepackage[cp1250]{inputenc}
\usepackage{txfonts}
\usepackage{bm}
\usepackage{graphicx}
\usepackage{natbib}
\bibpunct{(}{)}{;}{a}{}{,} 

\begin{document}
\title{Radio fiber bursts and fast magnetoacoustic wave trains}
\titlerunning{Fiber bursts and fast magnetoacoustic waves}

\author{M. Karlick\'y\inst{1} \and H. M\'esz\'arosov\'a\inst{1} \and P. Jel\'\i nek\inst{2}}
\offprints{M.~Karlick\'y, \email{karlicky@asu.cas.cz} } \institute{Astronomical
Institute of the Academy of Sciences of the Czech Republic,
           CZ--25165 Ond\v{r}ejov, Czech Republic
           \and
           University of South Bohemia, Faculty of Science, Institute of Physics and Biophysics, Brani\v sovsk\'a 10, CZ--370 05 \v{C}esk\'e
              Bud\v{e}jovice, Czech Republic}

\date{Received ................ / Accepted ...................}

\abstract
{}
{We present a model for dm-fiber bursts that is based on assuming fast sausage
magnetoacoustic wave trains that propagate along a dense vertical filament or
current sheet.}
{Eight groups of dm-fiber bursts that were observed during solar flares were
selected and analyzed by the wavelet analysis method. To model these fiber
bursts we built a semi-empirical model. We also did magnetohydrodynamic
simulations of a~propagation of the magnetoacoustic wave train in a~vertical
and gravitationally stratified current sheet.}
{In the wavelet spectra of the fiber bursts computed at different radio
frequencies we found the wavelet tadpoles, whose head maxima have the same
frequency drift as the drift of fiber bursts. It indicates that the drift of
these fiber bursts can be explained by the propagating fast sausage
magnetoacoustic wave train. Using new semi-empirical and magnetohydrodynamic
models with a~simple radio emission model we generated the artificial radio
spectra of the fiber bursts, which are similar to the observed ones.}
{}

\maketitle

\section{Introduction}
\begin{figure*}[ht]
\begin{center}
 \epsfig{file=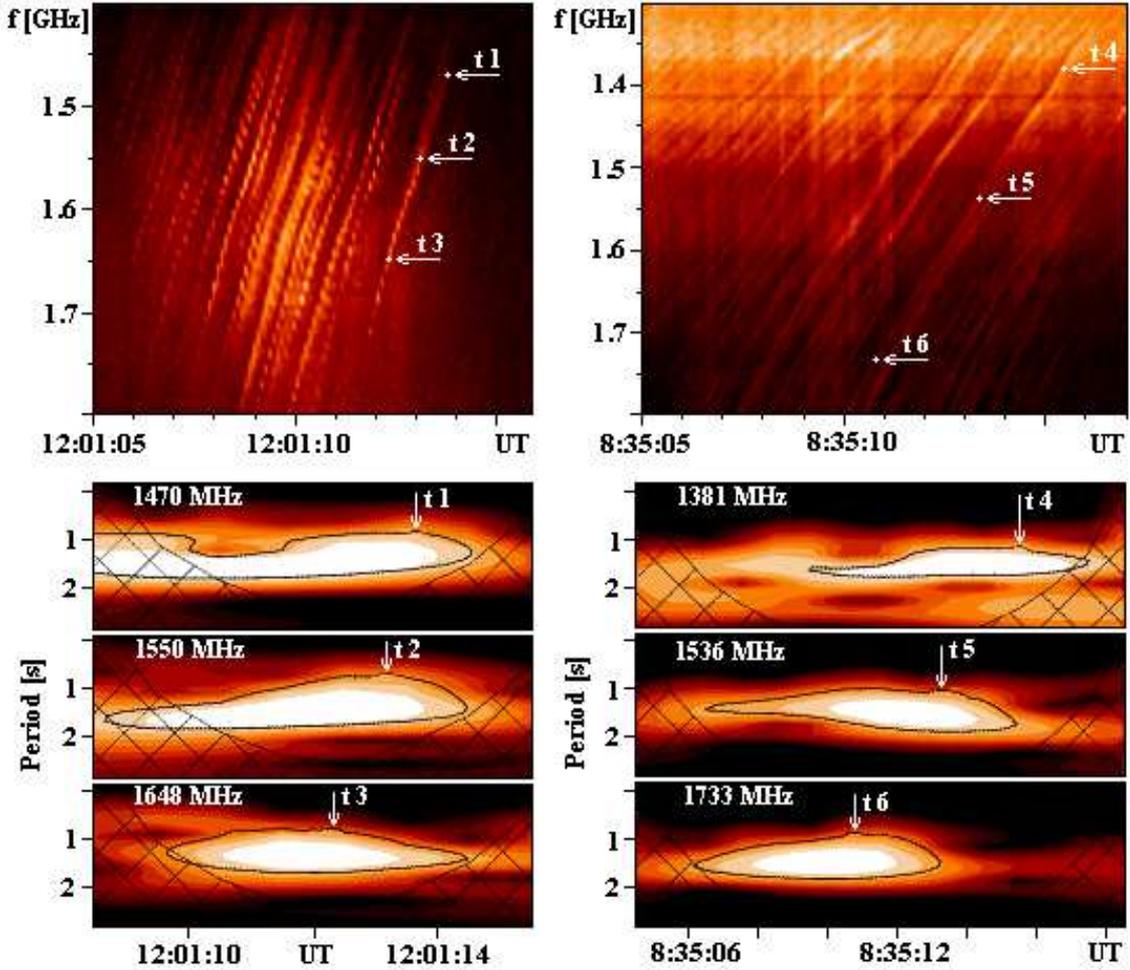, width=15cm}
\caption{Upper panels: Examples of the fiber bursts at
         12:01:05--12:01:16\,UT (November~23,~1998, left panel) and 8:35:05--8:35:17\,UT
         (November~18,~2003, right panel).
         Bottom part: Corresponding wavelet power spectra
         showing the tadpoles with the period $P$~=~1.4\,s (Table~2).
         In both spectra, at selected frequencies the times of the tadpoles'
         head maxima were determined ($t$1--$t$6) and also shown in the upper dynamic
         radio spectra (upper panels). The tadpole head maxima drift as the fiber bursts.}
\label{fig1}
\end{center}
\end{figure*}

Fiber bursts are considered to be a~fine structure of the broadband type~IV
radio bursts. In the dynamic spectrum, they occur in groups of many single
fibers, which are narrowband and have almost the same frequency drift. Because
this frequency drift is between those of types~II and III bursts, the fiber
bursts are also called the intermediate-drift bursts (Bernold \& Treumann 1983,
Aurass et al. 1987, Benz \& Mann 1998, Ji\v{r}i\v{c}ka et al. 2001).

There are two types of models of these bursts: a)~one based on the whistler
waves (Kuijpers 1975, Chernov 1976, Mann et al. 1987) and b)~one based on the
Alfv\'en waves or solitons (Treumann et al. 1990). Both these types of models
can be used for estimating the coronal magnetic fields (Benz \& Mann 1998,
Rausche et al. 2007), which is otherwise very difficult to determine. For this
reason the theory and interpretation of fiber bursts are very important.
However, these two types of models give different values for the magnetic field
strength, and there is still a~debate about which models are correct.

Besides these models, Kuznetsov (2006) proposes a model in which the fiber
bursts are generated by a~modulation of the radio emission by
magnetohydrodynamic waves. He also proposes that these waves could be
magnetoacoustic waves of a sausage mode type that propagates along the dense
coronal loop.

It is known that the structures with higher plasma density (like dense loops or
current sheets) act as waveguides, where the fast magnetoacoustic waves are
trapped (Roberts et al. 1984). If these waves are impulsively triggered at some
location in such a~dense structure, then at some distance from the initiation
site along this structure, these propagating waves form the wave trains owing
to their dispersion properties. The trains exhibit three phases: (1)~periodic
phase (long-period spectral components arrive as the first ones at the
observation point), (2)~quasi-periodic phase (both long- and short-period
spectral components arrive and interact), and finally (3)~decay phase (Roberts
et al. 1983 and 1984). As shown by Nakariakov et al. (2004), wavelet spectra of
these trains correspond to the wavelet tadpoles (tail plus head). Detailed
numerical simulations of these magnetoacoustic (fast sausage) wave trains and
corresponding tadpoles in the density slab, as well as in the current sheet,
were performed by Jel\'inek \& Karlick\'y (2012).

M\'esz\'arosov\'a et al. (2009a,~b) analyzed several radio events. In the first
paper we found the wavelet tadpoles in the gyro-synchrotron radio burst, where
all the tadpoles were detected at the same time in the whole frequency range.
In second paper, analyzing the 2005~July~11 radio burst (generated by the
plasma emission processes), we found the wavelet tadpoles that slowly drifted
with the frequency drift corresponding to the drift of the whole group of the
fiber bursts. Furthermore, M\'esz\'arosov\'a et al. (2011) repeated the wavelet
analysis of the 2005~July~11 radio event, considering a much broader range of
time periods, as well as longer time intervals than in the previous paper.
Thus, we found the tadpoles with the period of $P\approx$~1.9\,s and with the
fast frequency drift corresponding to the drift of individual fiber bursts.
Based on this, we proposed that the fiber bursts are generated by a modulation
of the type~IV radio emission by the magnetoacoustic wave trains. Karlick\'y et
al. (2011) also found the wavelet tadpoles in sources of the narrowband
dm-spikes.

In the present paper, we firstly verify the above-mentioned statement about the
similarity of the frequency drifts of individual fiber bursts to head maxima of
the corresponding tadpoles by the wavelet analysis of more observed groups of
fibers. Then we simulate fibers in a~newly developed semi-empirical model. Due
to limitations of this model we also made a first attempt to simulate the fiber
bursts in a~full magnetohydrodynamic (MHD) model with the fast sausage
magnetoacoustic wave train propagating in a~vertical and gravitationally
stratified current sheet. Using a simple radio emission model we produce
artificial radio spectra of the fiber bursts, which are similar to those
observed.

The paper is organized as follows: Section~2 describes the set of observed
fiber bursts and their wavelet analysis. In Section~3, we present our
semi-empirical model of fiber bursts and their simulations. Section~4 shows a
simulation of the fiber bursts using the MHD model. Finally, in Section~5 the
results are discussed.

\section{Fiber bursts observation and its waletet analysis}
We selected eight decimetric radio events with the fiber bursts recorded during
the years 1998--2005 by the Ond\v{r}ejov radiospectrograph (Ji\v{r}i\v{c}ka et
al. 1993). The times of their observations, the GOES X-ray and H$\alpha$
characteristics of associated flares are presented in Table~1. Examples of the
characteristic radio dynamic spectra that show two typical groups of fiber
bursts are presented in upper panels of Fig.~\ref{fig1} (1998~November~23,
12:01:05--12:01:16\,UT -- left panel, and 2003~November~18,
8:35:05--8:35:17\,UT -- right panel). The parameters of all fiber events are
summarized in Table~2. The fiber bursts were observed in the frequency range
1000--2000\,MHz, and their frequency drift $FD_{F}$ ranges from
$-73$\,MHz\,s$^{-1}$ to $-154$ \,MHz\,s$^{-1}$.

\begin{table*}[ht]
\begin{center}
\caption{Basic characteristics of the selected events with fiber bursts.}
\label{tab1}
\begin{tabular}{cc|cc|cccc|ccc}
\hline
\hline
    &                & \multicolumn{2}{|c|}{Radio} & \multicolumn{4}{c|}{GOES}   & \multicolumn{3}{c}{H$\alpha$} \\
No. & Flare          & Start  & End                & Start & Max  & End  & X-ray & Optic. & Position & NOAA      \\
    &                & [UT]   & [UT]               & [UT]  & [UT] & [UT] & Imp.  & Imp.   &          & AR \#     \\
\hline
1   & Nov. 23, 1998a & 11:51  & 11:52              & 10:59 &11:21 &13:25 & M3.1  &  1N    & S23 E58  & 8392      \\
2   & Nov. 23, 1998b & 11:58  & 11:59              & 10:59 &11:21 &13:25 & M3.1  &  1N    & S23 E58  & 8392      \\
3   & Nov. 23, 1998c & 12:00  & 12:02              & 10:59 &11:21 &13:25 & M3.1  &  1N    & S23 E58  & 8392      \\
4   & Mar. 05, 2000  & 10:03  & 10:04              &       &      &      & C6.8  &  SF    & S13 E43  & 8898      \\
5   & Apr. 26, 2001a & 11:30  & 11:31              & 11:26 &13:12 &13:19 & M7.8  &  2B    & N17 W31  & 9433      \\
6   & Apr. 26, 2001b & 11:41  & 11:44              & 11:26 &13:12 &13:19 & M7.8  &  2B    & N17 W31  & 9433      \\
7   & Nov. 18, 2003  & 08:34  & 08:36              & 08:12 &08:31 &08:59 & M3.9  &  1F    & N00 E19  &10501      \\
8   & Jul. 11, 2005  & 16:31  & 16:43              & 16:30 &16:38 &16:50 & C1.1  &  SF    & N09 W52  &10786      \\
\hline
\end{tabular}
\end{center}
\end{table*}

\begin{table*}[ht]
\begin{center}
\caption{Parameters of the fiber bursts and associated wavelet tadpoles.}
\label{tab2}
\begin{tabular}{c|cc|cc}
\hline
\hline
    &     \multicolumn{2}{|c}{Fiber bursts}       &  \multicolumn{2}{|c}{Wavelet tadpoles}  \\
No. & Frequency range & $FD_{F}$     &  Period &  $FD_{T}$    \\
    &    [MHz]        &  [MHz~s$^{-1}$]           &    [s]  & [MHz~s$^{-1}$]                \\
\hline
 1  & 1456 -- 2000      &  $-$154                     &    1.5  &  $-$152             \\
 2  & 1370 -- 1800      &  $-$130                     &    1.5  &  $-$122             \\
 3  & 1370 -- 1800      &  $-$128                     &    1.4  &  $-$127             \\
 4  & 1000 -- 1330      &  $-$144                     &    1.1  &  $-$136             \\
 5  & 1050 -- 1600      &  $-$92                     &    2.0  &   $-$87             \\
 6  & 1000 -- 1600      &  $-$76                     &    2.3  &   $-$61             \\
 7  & 1100 -- 2000      &  $-$73                     &    1.4  &   $-$75             \\
 8  & 1100 -- 1800      &  $-$78                     &    1.9  &   $-$86             \\
\hline
\end{tabular}
\end{center}
\end{table*}

The selected groups of fibers (Table~1) were analyzed using the wavelet method
as described by M\'esz\'arosov\'a et al. (2009a,~b). We searched for tadpoles
in the wavelet power spectra of radio flux time series and selected only
dominant tadpoles corresponding to the 99\% confidence level. The confidence
level implies a~test against a~certain background level. If a~peak in the
wavelet power spectrum is significantly above background spectrum, then it can
be assumed to be a~true feature with a~certain percent of the confidence (Farge
1992). In the present paper the background spectrum is modeled by the red
noise, and a~computation of the confidence level is performed as described by
Torrence \& Compo (1998). The detected wavelet tadpoles and their parameters
are summarized in Table~2. Their period~$P$ ranges from 1.1 to 2.3\,s, and the
drift of their head maxima $FD_{T}$ is in the range from $-$61\,MHz\,s$^{-1}$
to $-$152\,MHz\,s$^{-1}$. Comparing the frequency drifts of all fiber bursts in
Table~2, we can see that the drift of fiber bursts is very similar to that of
the head maxima of the corresponding tadpoles (compare values in columns~3
and~5 in Table~2). This can be also seen in Fig.~\ref{fig1}: see the times
$t$1~=~12:01:13.7, $t$2~=~12:01:13.1, $t$3~=~12:01:12.3, $t$4~=~8:35:15.5,
$t$5~=~8:35:13.3, $t$6~=~8:35:10.8\,UT at the selected frequencies 1470, 1550,
1648, 1381, 1536, and 1733\,MHz, respectively.

Because the wavelet tadpoles are interpreted as signatures of the fast sausage
magnetoacoustic wave trains (e.g. Nakariakov et al. 2004), the similarity of
both the drifts indicates that the fiber bursts are physically connected with
these magnetoacoustic wave trains. Based on these results we propose that the
analyzed dm-fiber bursts are generated by the fast sausage magnetoacoustic wave
which modulates the radio emission of superthermal electrons trapped in a~flare
loop or in current sheet.

\section{Semi-empirical model}
This model is similar to the one presented by Kuznetsov (2006). However, it
differs in one aspect: for the fast sausage magnetoacoustic waves considered
here, the magnetic and density wave perturbations ($b_z$~and~$\rho_1$) are in
phase (Erdelyi, priv. communication, 2012), contrary to Kuznetsov (2006), where
these perturbations are in anti-phase.

\begin{figure}[ht]
\begin{center}
  \epsfig{file=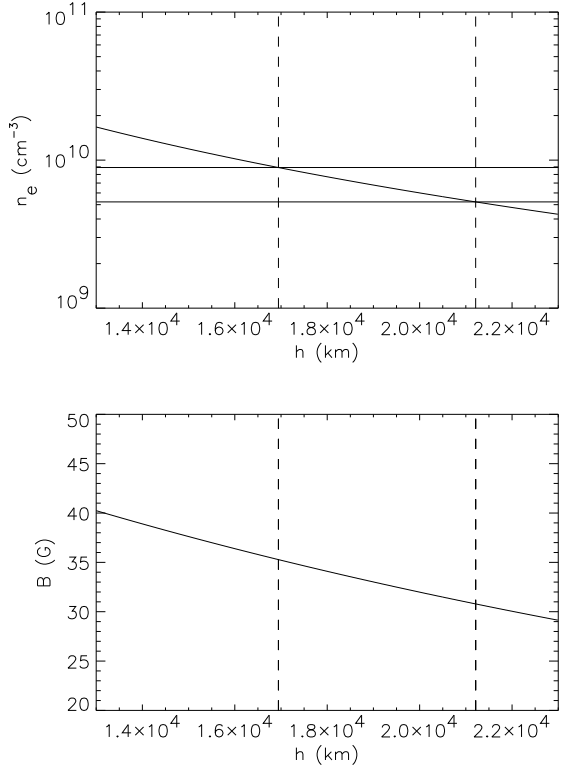, width=8cm}
  \caption{Plasma density and magnetic field profiles in dependence on the height
  in the modeled loop. The horizontal full lines in the density plot mean
  the plasma densities corresponding to the maximum and minimum of
  frequencies considered in the radio spectra. The vertical dashed lines limit
  this density interval.}
  \label{fig2}
\end{center}
\end{figure}

We assume that the magnetoacoustic wave is triggered by a~flare and propagates
upwards along a~dense vertical flare loop. The plasma density and magnetic
field in the loop are taken according to the models by Aschwanden (2002, 2004),
where the magnetic field is expressed as $B(h)~=~B_{00}~(1+h/h_D)^{-3}$ ($h$~is
the height in the solar atmosphere, $h_D$~=~75\,Mm, and $B_{00}$ is the
footpoint magnetic field taken in the following runs No.~1 and~2 as
$B_{00}$~=~65\,G). The plasma density and magnetic field profiles used in the
model are shown in Fig.~\ref{fig2}.

Although a~real wave evolution is complex (see Roberts et al. (1984),
Nakariakov et al. (2004), and also the following MHD model), for
a~simplification we take the velocity of this wave perturbation as the Alfv\'en
speed~$v_A$ computed along the loop. The ratios of wave perturbations is taken
as $b_z/B_0$~=~$\rho_1/\rho_0 \ll$~1, where $B_0$~and~$\rho_0$ mean the local
magnetic field and density in the Aschwanden's models.

We took a~wave in the form of the wave packet as
\begin{eqnarray}
F(h, t) = F_0 C_0 \exp{\left[-\left(\frac{h - v_A(h) t -
h_0}{d_s}\right)^2\right]} \times
\nonumber \\
\times \cos{\left[\frac{2 \pi (h - v_A(h) t - h_0)}{d_L}\right]},
\end{eqnarray}
\begin{eqnarray}
C_0 = \exp{\left[-\left(\frac{t - t_{Amax}}{d_t}\right)^2\right]},\nonumber \\
\end{eqnarray}
where $F(h,t)$~means the density~$\rho_1$ or magnetic field~$b_z$ wave
perturbation, $F_0$~is the wave amplitude, $h$ the height in the loop, $t$ the
time, $v_A$ the local Alfv\'en speed, $h_0$ the wave initiation height, $d_s$
the spatial width of the wave packet, and $d_L$ the wavelength. By
changing~$d_s$ for a~fixed $d_L$~we can generate the wave for a~single fiber or
group of fibers. We also added a~factor~$C_0$ that limits radio emission to our
chosen frequency interval 1300--1700\,MHz (see the following): $t_{Amax}$~is
thus the time of the maximum wave amplitude, and $d_t$ the corresponding
characteristic time. In our computations the time $t$~changes from~0 to~5\,s,
$t_{Amax}$~=~2.5\,s, $d_t$~=~1.5\,s, and $h_0$~=~16.9\,Mm. The other parameters
are presented in Table~3. This propagating wave packet is then superimposed on
the density and magnetic field profiles presented in Fig.~\ref{fig2}.

\begin{figure}[ht]
\begin{center}
  \epsfig{file=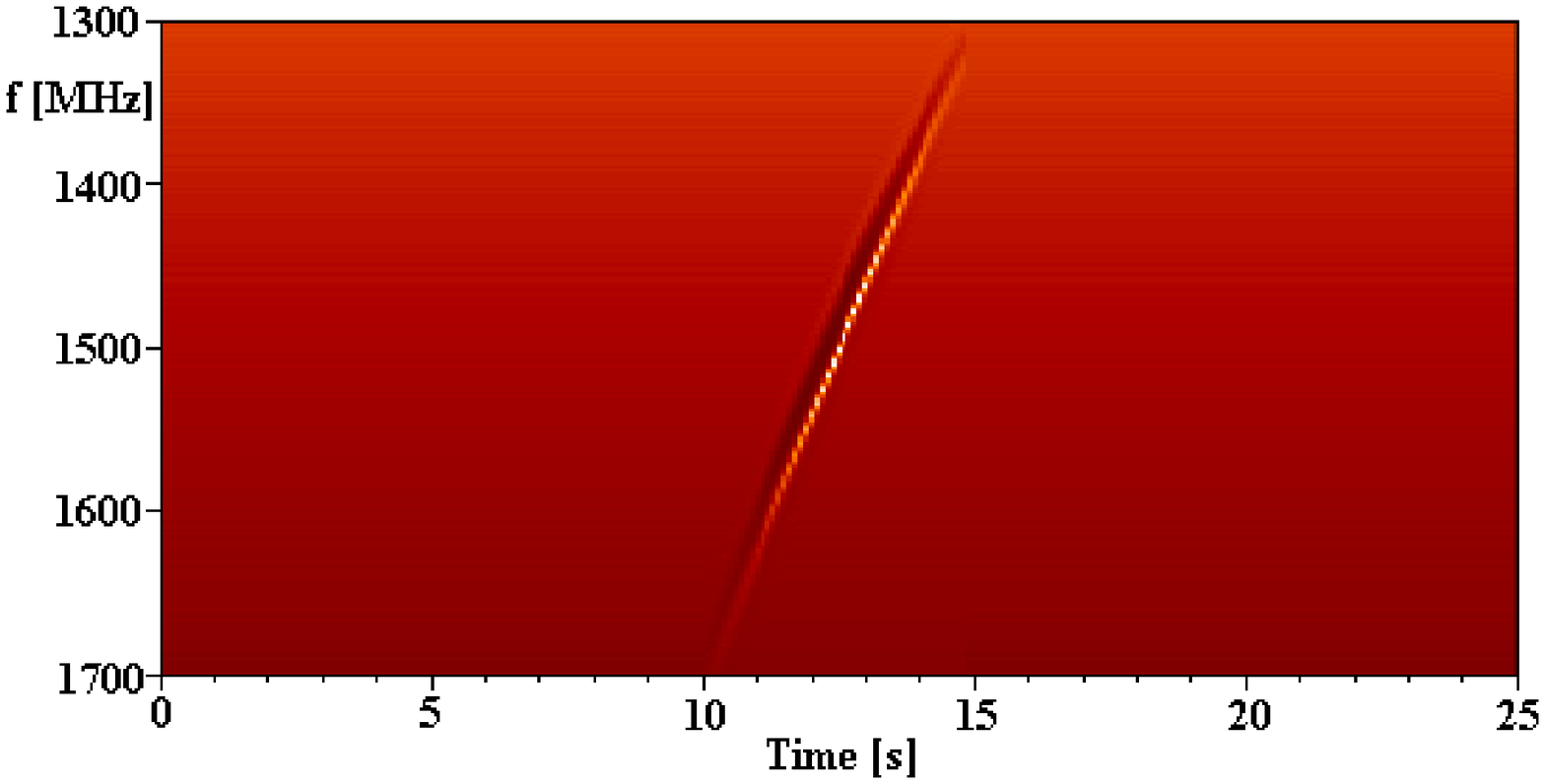, width=9cm}
  \caption{Computed radio spectrum with the fiber modeled in the semi-empirical
  model for run No.~1.}
  \label{fig3}
\end{center}
\end{figure}

\begin{figure}[ht]
\begin{center}
  \epsfig{file=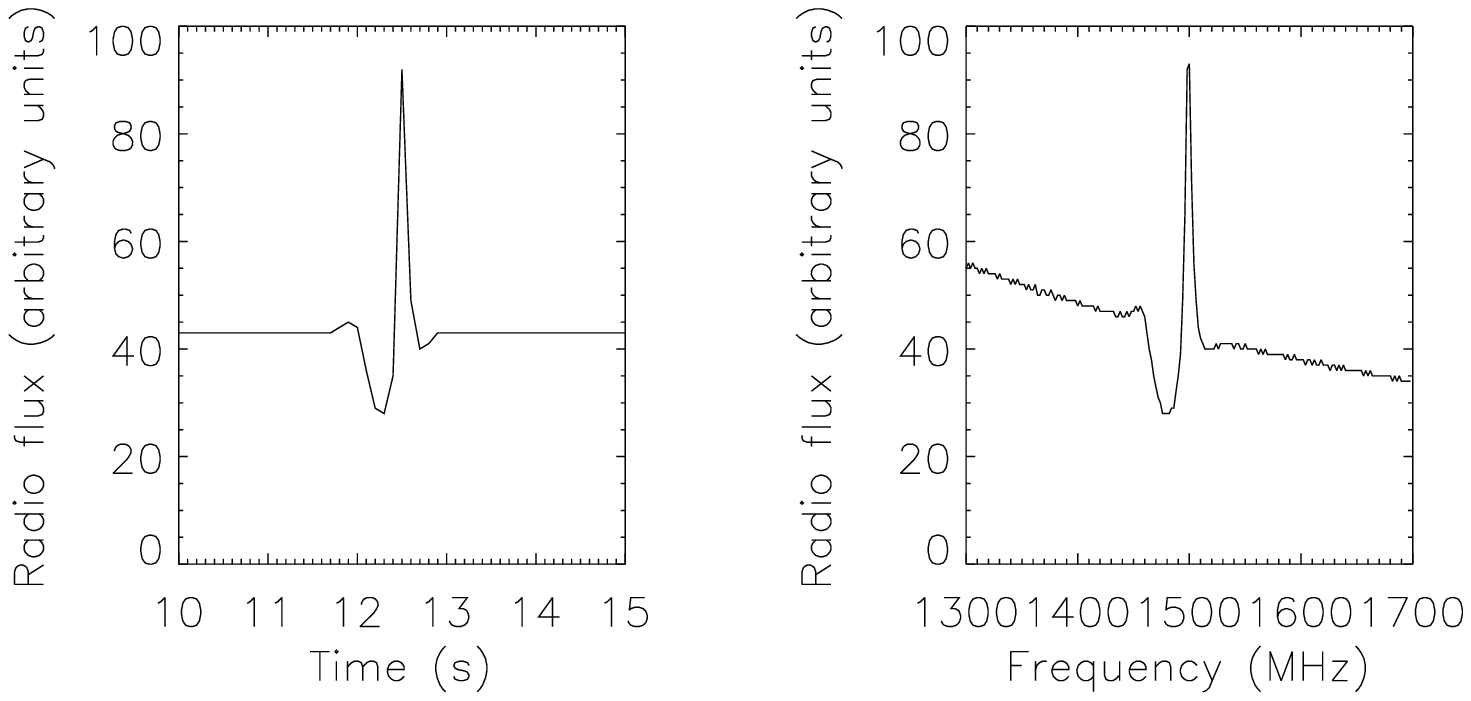, width=8cm}
  \caption{Example of the time and frequency profiles of the fiber burst
  modeled in the semi-empirical model for run No.~1. The time profile is plotted at the 1500\,MHz frequency
  and the frequency profile at the time 12.5\,s; compare with the spectrum in Fig.~3.}
  \label{fig4}
\end{center}
\end{figure}

\begin{figure}[ht]
\begin{center}
  \epsfig{file=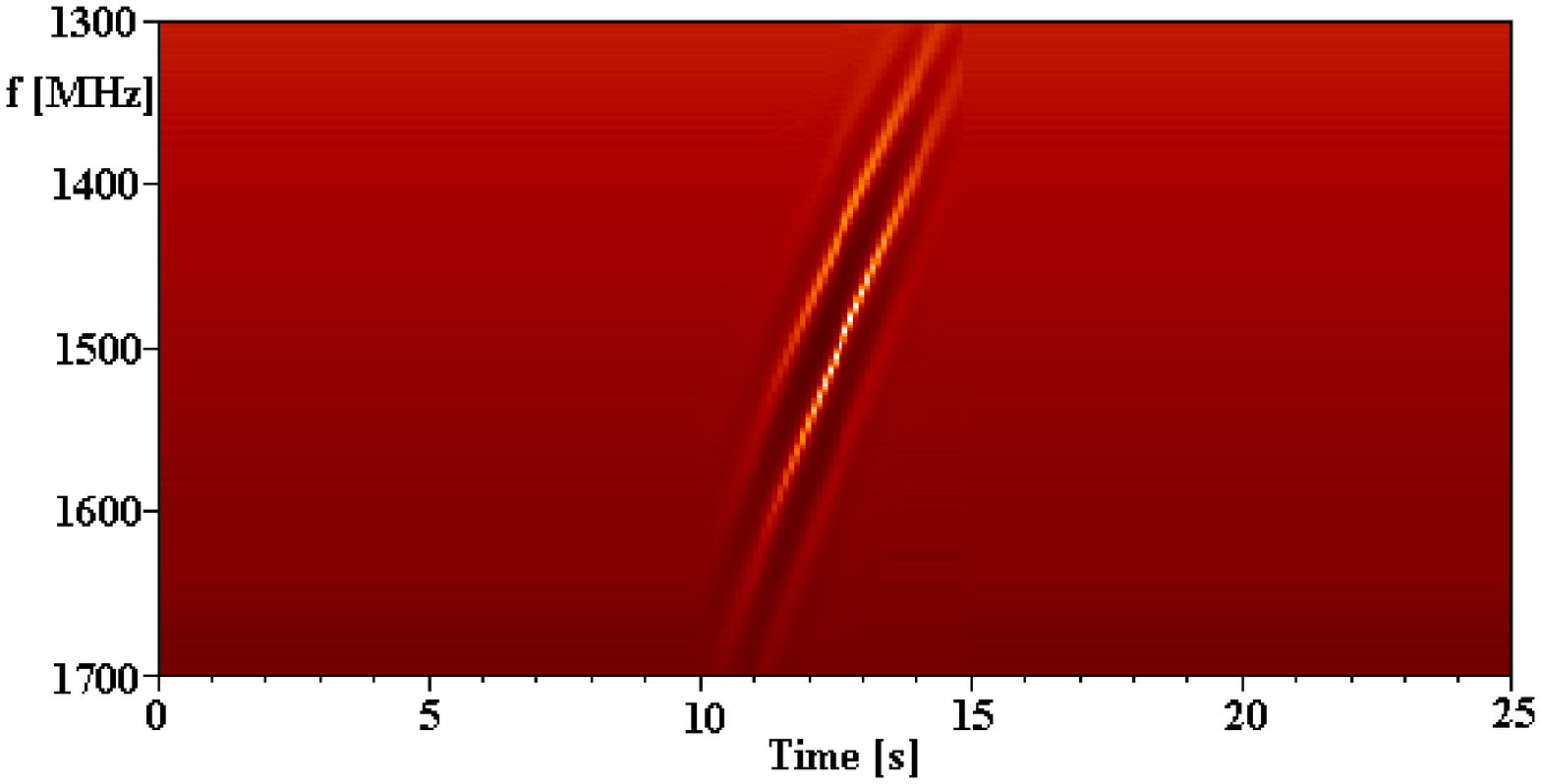, width=9cm}
  \caption{Computed radio spectrum with the fibers modeled in the semi-empirical
  model for run No.~2.}
  \label{fig5}
\end{center}
\end{figure}

\begin{figure}[ht]
\begin{center}
  \epsfig{file=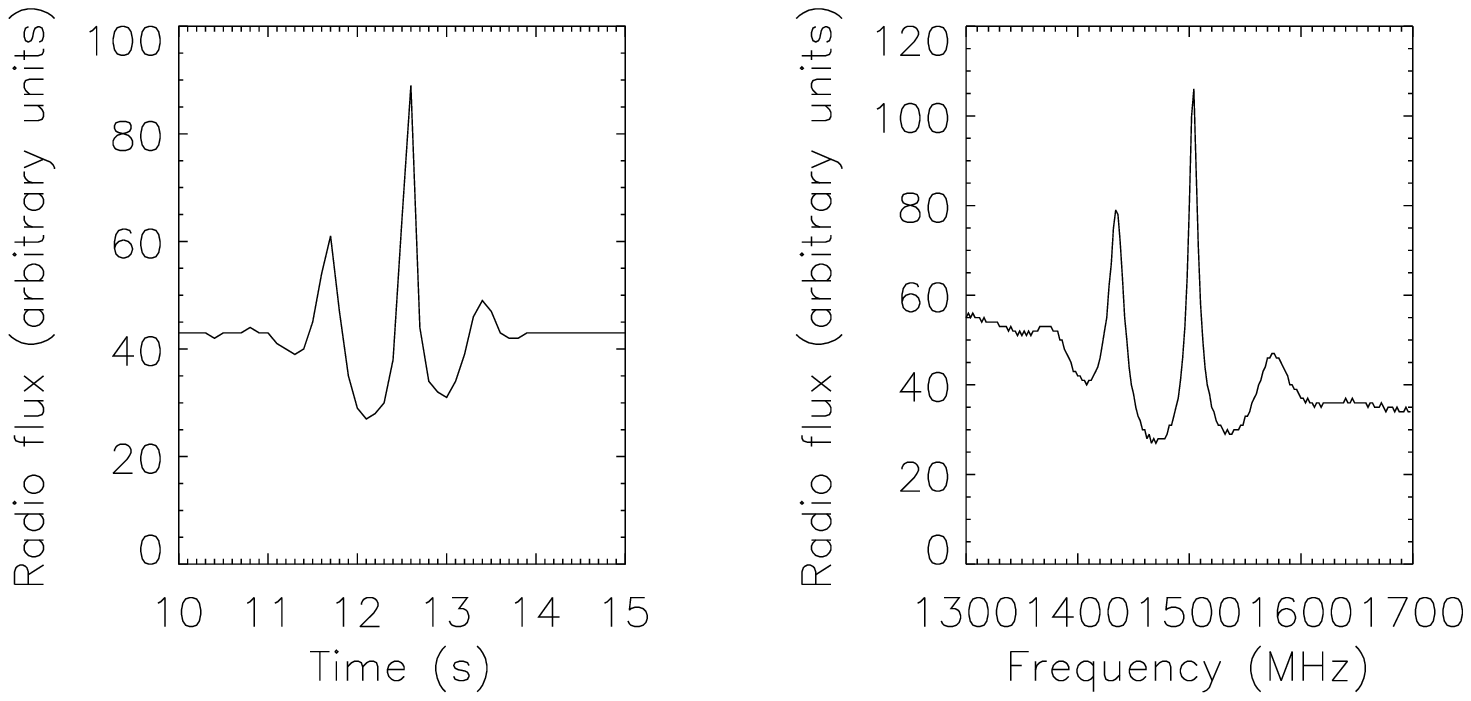, width=8cm}
  \caption{Example of the time and frequency profiles of the fiber bursts
  modeled in the semi-empirical model for the run No.~2. The time profile is plotted
  at 1500\,MHz frequency and the frequency profile at time 12.5\,s;
  compare with the spectrum in Fig.~5.}
  \label{fig6}
\end{center}
\end{figure}

\begin{figure}[ht]
\begin{center}
  \epsfig{file=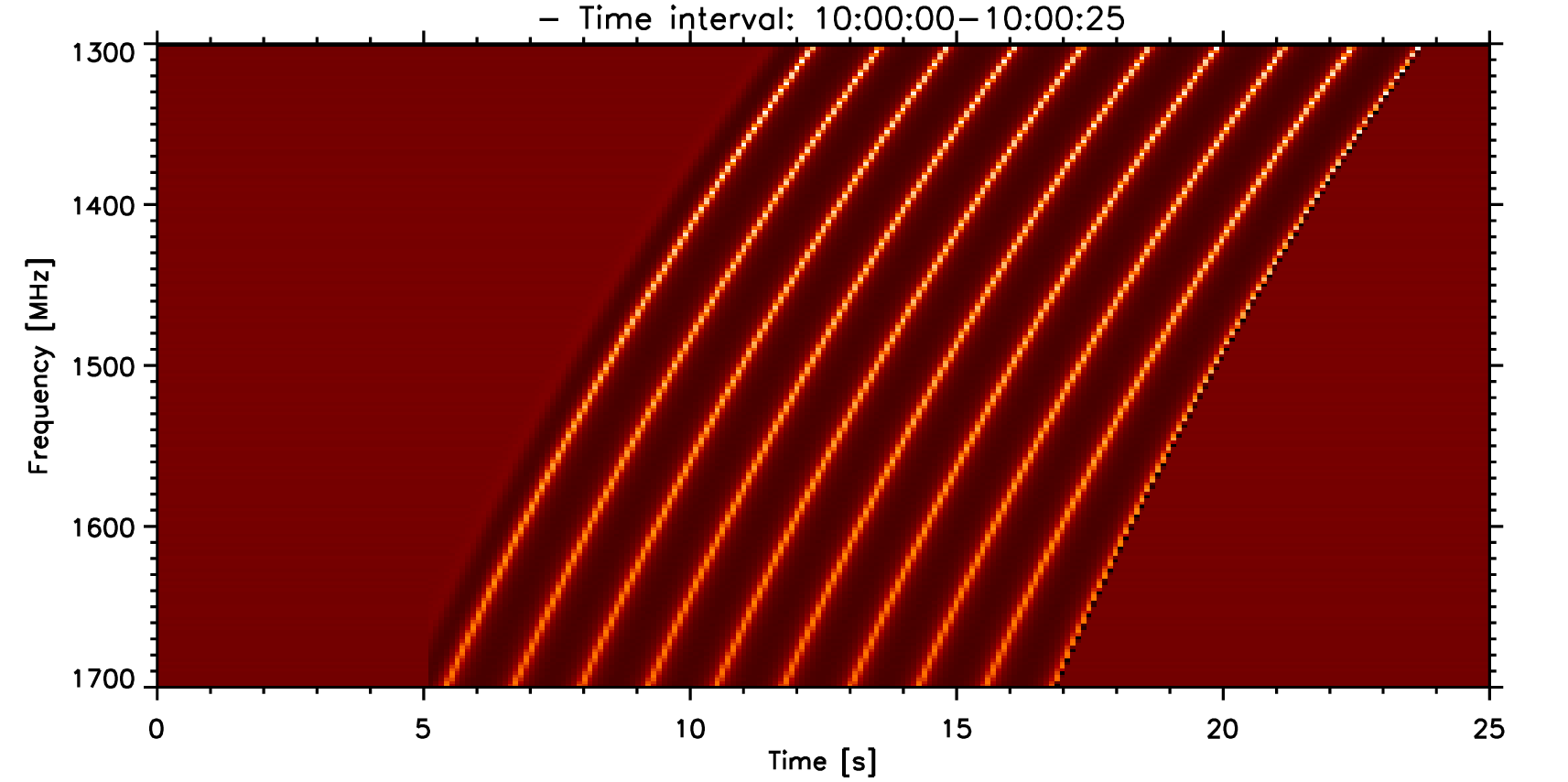, width=4.5cm, angle=90}
  \caption{Computed radio spectrum with the 10 fiber bursts modeled
  by the semi-empirical model.}
  \label{fig7}
\end{center}
\end{figure}

In agreement with Kuznetsov (2006), we assume that the radio emission is
produced at the double (harmonic) upper hybrid frequency as a~result of
a~coalescence of two upper hybrid waves, generated by the loss-cone instability
of superthermal electrons. Furthermore, we assume that the radio emission is
generated with a~constant intensity in the whole interval of considered
heights. It means that every unit of the volume along the flaring loop (or
current sheet) generates the same radio emission. Thus, the intensity of the
radio emission at some specific frequency depends only on the gradients of
$n_e$~and~$B$ in the radio source; i.e., the wave perturbation modulates the
radio emission as proposed by Treumann et al. (1990). Then, the resulting radio
flux at the specific frequency~$f$ can be expressed as a~sum of emissions from
plasma volumes having the upper hybrid frequency~$f_{up}$ close to~$f/2$.

\begin{table}[h!]
\caption{Computation parameters.}
\label{tab4}
\centering
\begin{tabular}{ccccc}
\hline\hline
Run No. & $\rho_1/\rho_0$ & $b_z/B_0$ & $d_s$ (km) & $d_L$ (km) \\
\hline
1 & 0.01 & 0.01 & 200 & 800 \\
2 & 0.01 & 0.01 & 800 & 800 \\
\hline
\end{tabular}
\end{table}

First, we made two runs with the parameters shown in Table~3. The first run was
performed for a~single perturbation pulse and the second one for the wave
train. The computed radio spectrum is in Fig.~\ref{fig3}. (Remark: For better
presentation of all computed spectra at their beginning we added the 10-second
radio spectrum of the unperturbed state.) An example of the time and frequency
profiles of the resulting fiber for run No.~1 (cuts in the spectrum shown in
Fig.~\ref{fig3}) is presented in Fig.~\ref{fig4}. Besides the emission peak the
emission depression on the low-frequency side of the fiber, as observed, is
also clearly visible. The radio spectrum and time and frequency profiles for
run No.~2, i.e. for the magnetoacoustic wave train, are shown in
Figs.~\ref{fig5} and~\ref{fig6}. As can be seen here, the wave train generates
a~series of fiber bursts.

Furthermore, as an example, we simulated a~group of ten fiber bursts. We
consider the wave which is spatially limited to ten wavelengths (wavelength
$d_L$~=~800\,km). It propagates upwards in the solar atmosphere as in runs~1
and~2. However, the magnetic field~$B_{00}$ in the used magnetic-field model is
$B_{00}$~=~48\,G. The constant~$C_0$ (see Eq.~2) is taken as $C_0$~=~1, which
means no global frequency modulation of the radio flux. The computed spectrum
with ten fiber bursts is shown in Fig.~\ref{fig7}.

Comparing this figure with the radio spectrum in the righthand panel of
Fig.~\ref{fig1}, we can see some similarities as well as some deviations.
Generally, using this semi-empirical model, we can simulate the frequency drift
and repetition period of the fiber bursts by varying the Alfv\'en speed
(through the magnetic field~$B_{00}$) and wavelength of the magnetoacoustic
wave ($d_L$). On the other hand, deviations are caused by limited information
about conditions in the real radio source and also by limitations in the
semi-empirical model used. One of the main problem of this semi-empirical model
is that dispersive properties of the magnetoacoustic waves are neglected.
Therefore, in the following we describe a~first attempt to simulate the fiber
bursts in a~full MHD model, where these properties are implicitly included.

\section{MHD model}
The main problem of any such model is an initiation of the sufficiently good
waveguide for the fast sausage magnetoacoustic waves, which is in equilibrium
in the gravitationally stratified solar atmosphere. For a~waveguide we can use
the dense loop or current sheet, see Jel\'inek \& Karlick\'y (2012). In an
analytical form, the model of the vertical current sheet in the gravitationally
stratified atmosphere proposed by Galsgaard and Roussev (2002) is an
appropriate one for initiating the equilibrium waveguide.

In our case the vertical current sheet (it would be better to call it the
current layer) is only a~prototype of the waveguide guiding the magnetoacoustic
waves. We did not consider any reconnection and fragmentation processes in this
current sheet as discussed in B\'arta et al. (2011a,~b), Karlick\'y et al.
(2012), and Cargill et al. (2012), among others. Although the fragmentation is
real in these current sheets, we think that such vertical current sheets can
exist; see, e.g., the observed current sheet and its width on the order of
1000\,km in Fig.~3 in the paper by Lin et al. (2005).

In our model we use the ideal 2D magnetohydrodynamic (MHD) equations:
\begin{equation}\label{eq1}
\frac{\mathrm{D}\varrho}{\mathrm{D}t} = -\varrho \nabla \bm{v},
\end{equation}
\begin{equation}\label{eq2}
\varrho \frac{\mathrm{D}\bm{v}}{\mathrm{D}t} = -\nabla p+\bm{j}\times\bm{B} +
\varrho \bm{g},
\end{equation}
\begin{equation}\label{eq3}
\frac{\mathrm{D}\bm{B}}{\mathrm{D}t} = (\bm{B} \cdot \nabla)\bm{v},
\end{equation}
\begin{equation}\label{eq4}
\frac{\mathrm{D}e}{\mathrm{D}t} = -\gamma e \nabla \cdot \bm{v},
\end{equation}
\begin{equation}\label{eq5}
\nabla\cdot\bm{B}=0,
\end{equation}
where $\mathrm{D} / \mathrm{D}t \equiv \partial / \partial t + \bm{v} \cdot
\nabla$ is the total time derivative, $\varrho$~is a~mass density, $\bm{v}$
flow velocity, $\bm{B}$ the magnetic field, and $\bm{g}=[0,-g_{\sun},0]$ is the
gravitational acceleration with $g_{\sun}~=~274~\mathrm{m \cdot s^{-2}}$. The
current density~$\bm{j}$ in Eq.~(\ref{eq2}) is expressed as
\begin{equation}\label{eq6}
\bm{j} = \frac{1}{\mu_0}\nabla \times \bm{B},
\end{equation}
where $\mu_0 = 1.26 \times 10^{-6}~\mathrm{H\cdot m}^{-1}$ is the magnetic
permeability of free space. The specific internal energy~$e$ in Eq.~(\ref{eq4}) is given by
\begin{equation}\label{eq7}
e = \frac{p}{(\gamma - 1)\varrho},
\end{equation}
with the adiabatic coefficient~$\gamma = 5/3$.

Equations (\ref{eq1})-(\ref{eq4}) are solved numerically by the FLASH code
(Fryxell et al. 2000, Lee \& Deane 2009). This code implements second- and
third-order unsplit Godunov solvers and adaptive mesh refinement (AMR) (see
e.g. Chung 2002, Murawski 2002). Spatial resolution of the numerical grid is
determined by the AMR method, and we use the AMR grid with a~minimum (maximum)
level of refinement blocks set to 3 (6). The total number of computational
cells covering a~whole simulation plane is 267456. A~spatial cell size has to
be much smaller than the width of the current sheet in the x-direction and the
minimal wavelength of the magnetoacoustic waves in the h-direction. In our
model the minimal cell sizes are $\Delta x$~=~0.0175\,Mm and $\Delta
h$~=~0.005\,Mm (compared with the half width of the current sheet
$w_\mathrm{CS}$~=~0.35\,Mm and the minimal wavelength of about 1.0\,Mm).

At the start of our numerical calculations we set a~refinement procedure in the
region covering the current sheet. During the calculations, the FLASH code
automatically controlled in each time step the gradient of mass density, which
value is then used for refining the grids. For our numerical simulations we
used the two-dimensional (2-D) Eulerian box $(-2.25,2.25)~\mathrm{Mm} \times
(13.0,23.0)~\mathrm{Mm}$ in the~$x$- and $h$-~(height) directions,
respectively. As a~consequence of an extension of the real plasma medium we
apply free-boundary conditions at the boundaries of the simulation region, so
that the waves can freely leave the simulation box without any significant
reflection.

For the description of magnetic field in the vertical current sheet in
gravitationally stratified solar atmosphere in the $x-h$ plane we use following
expressions:
\begin{equation}\label{eq15}
B_x(x,h) = B_{\mathrm{o}} \frac{w_\mathrm{cs}}{\lambda} \ln \left[\cosh
\left(\frac{x}{w_\mathrm{cs}}\right)\right] \exp
\left(-\frac{h}{\lambda}\right),
\end{equation}

\begin{equation}\label{eq16}
B_h(x,h) = B_{\mathrm{o}} \tanh \left(\frac{x}{w_\mathrm{cs}}\right)\exp
\left(-\frac{h}{\lambda}\right),
\end{equation}
where $B_\mathrm{o}$~is a~magnetic field at $x \rightarrow \infty$,
$w_\mathrm{CS}$~is the half width of the current sheet, and $\lambda$ denotes
the magnetic scale-height. For an equilibrium ($\bm{v} = \bm{0}$), the Lorentz
and gravity forces have to be balanced by the pressure gradient in the entire
physical domain
\begin{equation}\label{eq8}
-\nabla p+\bm{j}\times\bm{B} + \varrho \bm{g} = \bm{0}.
\end{equation}
From this condition we can derive the formulae for distribution of the mass
density and gas pressure. For the details see Galsgaard \& Roussev (2002) and
Jel\'inek et al. (2012).

In the initial state, we generate the vertical current sheet with the density
profile along its axis according to the Aschwanden model (2002). The half width
of the current sheet at the height $h$~=~16.9\,Mm is taken as
$w_\mathrm{CS}~=~0.35~\mathrm{Mm}$. In the paper by Jel\'inek \& Karlick\'y
(2012), we found that the neutral current sheet serves as the same waveguide as
the dense slab if the Alfv\'en speed at the half width of the current sheet is
the same as in the slab. Therefore, in the present current sheet we took this
Alfv\'en speed as 690\,km\,s$^{-1}$. The change of this speed in the studied
interval of heights in the current sheet is small.

At the start of the numerical simulation, the initial equilibrium state is
perturbed by the Gaussian pulse in the $x$-component of velocity and has the
following form (e.g. Nakariakov et al. 2004, 2005):

\begin{equation}\label{eq22}
v_x = -A_0\cdot \frac{x}{\lambda_h} \cdot
\exp{\left[-\frac{(x-h_{\mathrm{P}})^2}{\lambda_x^2}\right]}\cdot
\exp{\left[-\frac{h^2}{\lambda_h^2}\right]},
\end{equation}
where $A_0$~is the initial amplitude of the pulse, and $\lambda_x = \lambda_h =
0.35~\mathrm{Mm}$ are the widths of the velocity pulse in the longitudinal and
transverse directions. This pulse tends to trigger fast sausage magnetoacoustic
waves.  The initial perturbation is located at the height
$h_{\mathrm{P}}$~=~15\,Mm.

\begin{figure}[ht]
\begin{center}
  \epsfig{file=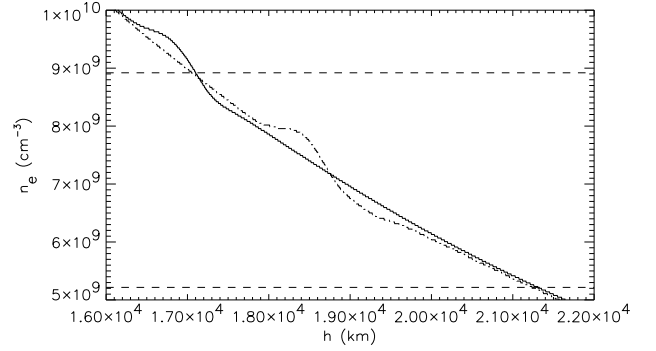, width=8cm}
  \caption{The density profiles in the MHD model at 11\,s (full line) and 14\,s (dash-dotted line).
  The times correspond to times in the spectrum in Fig.~9.}
  \label{fig8}
\end{center}
\end{figure}

\begin{figure}[ht]
\begin{center}
  \epsfig{file=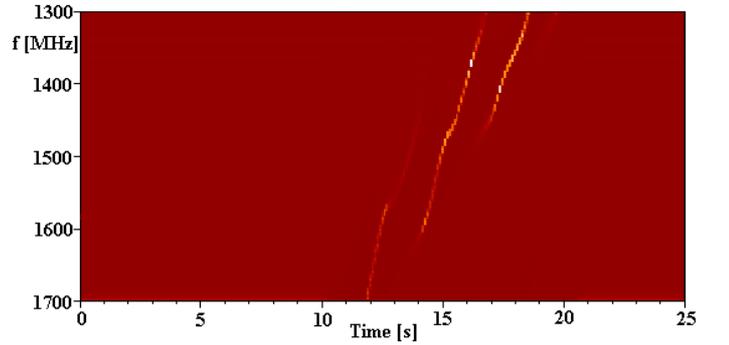, width=9cm}
  \caption{The computed radio spectrum with the fibers modeled in the MHD model.}
  \label{fig9}
\end{center}
\end{figure}

\begin{figure}[ht]
\begin{center}
  \epsfig{file=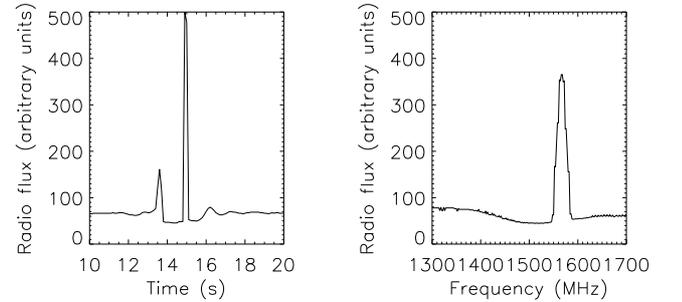, width=9cm}
  \caption{An example of the time and frequency profiles of the fiber bursts
  modeled in the MHD model. The time profile is plotted at the 1500\,MHz frequency
  and the frequency profile at the time 14.5\,s; compare with the spectrum in Fig.~9.}
  \label{fig10}
\end{center}
\end{figure}

Using this MHD model, we computed an evolution of the density profiles along
the axis of the vertical current sheet (see examples in Fig.~\ref{fig8}). These
density profiles were then used in computations of the artificial radio
spectrum, as well as the time and frequency profiles shown in Figs.~\ref{fig9}
and~\ref{fig10}. We used the same simple radio emission model as in the case of
the semi-empirical model. As seen in Fig.~\ref{fig9} the frequency drift of
these fiber bursts partly changes. It is caused by the wave dispersive effects
(included in this MHD model), which change the density wave profile during the
wave propagation. Compare this with the semi-empirical model, where the
perturbation profile is rigid. The maximum of fiber bursts are produced at the
locations with minimal density gradients and these gradients evolve in time.
Similarly, the emission depressions, corresponding to the maximal density
gradients, evolve in time.

The detailed analysis of these results shows that the model used for the
vertical current-sheet is not the best waveguide for the studied
magnetoacoustic waves. The change in the magnetic field gradient across the
current sheet is not sharp enough. Therefore, some wave energy escapes from
this waveguide. This wave energy leakage limits the wave train length and thus
also the number and duration of computed fibers. We hope that in a~future model
we will generate a~better waveguide, in which a longer wave train and thus more
fiber bursts will be generated by a~single perturbation. On the other hand,
more fiber bursts can be generated by a~time repetition of the initial
perturbation because this is also possible in real conditions. Furthermore,
some real radio spectra might show a~superposition of the fiber bursts
generated in several nearby waveguides.

\section{Discussion and conclusions}
After analyzing eight groups of observed fiber bursts by the wavelet technique,
we found a~connection between the dm-fiber bursts and fast sausage
magnetoacoustic waves. Based on this connection, we developed the model of the
dm-fiber bursts with the fast sausage magnetoacoustic wave propagating in the
waveguide of the enhanced plasma density: (a)~in the dense loop according to
the Aschwanden's density and magnetic field models (semi-empirical model) and
(b)~in the vertical and gravitationally stratified current-sheet (MHD) model.
In both of these models, using the simple emission model according to Treumann
et al. (1990), we computed the artificial radio spectra with the fiber bursts
that resembles to the observed ones.

We assumed that every unit of the volume along the flaring loop (or current
sheet) generates the same radio emission. The intensity of the radio emission
at some specific frequency depends only on the density gradient in the radio
source. However, the wave perturbation (with weak local magnetic mirrors) can
also modulate the growth rate of the upper hybrid waves through change in the
loss-cone angle of superthermal electrons (Winglee \& Dulk 1986, Yasnov \&
Karlick\'y 2004) and thus change local emissions. For this purpose, the
weighting function~$w_e (h(t))$ that expresses these variations can be included
in future models.

In all studied decimetric fiber bursts, the frequency drift of the wavelet
tadpoles corresponds to the drift of individual fiber bursts. It speaks in
favor of the present model with the magnetoacoustic waves in comparison to the
model with whistler waves, which is usually considered in the metric frequency
range. We propose to do similar wavelet analysis of the metric fiber bursts and
to check the validity of their models.

While in the semi-empirical model the form of the wave perturbation is rigid,
the MHD model describes perturbations including dispersive properties of the
magnetoacoustic waves. It leads to a~wavy character of the fiber bursts, which
can even be seen in some observed fiber bursts (see e.g. the fibers in the
upper-left part of Fig.~\ref{fig1}).

The semi-empirical model is very similar to that of Kuznetsov (2006), except
that the density and magnetic field perturbations are in phase. The phase
relation presented in Kuznetsov (2006) is not correct for the fast sausage
magnetoacoustic waves. However, this difference in both the models leads to
only small differences in the radio spectrum, because the electron-cyclotron
frequency~$\omega_{ce}$ is much lower than the plasma frequency~$\omega_{pe}$,
and thus the upper-hybrid frequency roughly equals the plasma frequency and the
magnetic field can be neglected in this case. However, the variation of the
magnetic field in the magnetoacoustic wave train can influence the growth rate
of the upper hybrid waves owing to the loss-cone instability of superthermal
electrons. It can modify the radio emission at some locations along the wave
train and thus change the fiber burst profiles. This effect is not considered
in the present paper.

Comparing both these models, the semi-empirical is easy to use, however, with
a~limited description of physical processes under study. The MHD model is much
more complicated. One of the main problems is to generate the appropriate
initial equilibrium state. The MHD model includes the dispersive effects of the
propagating magnetoacoustic waves, so it describes the fiber burst generation
in a~more realistic way. However, the present MHD vertical and gravitationally
stratified waveguide (the current sheet according to the Galsgaard and Roussev
model, 2002) is not ideal. Owing to its relatively smooth boundaries, the
energy of the magnetoacoustic waves escape from this wave guide. The energy of
wave decreases, and it is therefore difficult to make long-lasting wave train.
This means that we can only simulate a~few fiber bursts per one perturbation
instead of the long series as in the semi-empirical model. This problem is
connected to a~solution of the initial equilibrium waveguide. Therefore, to
make this MHD model of the fiber bursts more realistic, a~better initial
waveguide needs to be found. On the other hand, the series of fiber bursts can
be generated in the present MHD model using time series of perturbations.
Furthermore, some of real radio spectra might show a~superposition of the fiber
bursts generated in several nearby waveguides.

Considering our simple emission model, we plan to solve the inverse problem to
the presented solution: i.e., we plan to determine the density profiles of the
propagating magnetoacoustic wave from the fiber burst profiles measured along
the radio frequency at some specific times.

\begin{acknowledgements}
The authors thank an anonymous referee for comments that improved the paper.
This research was supported by grants P209/12/0103 (GA CR), P209/10/1680 (GA
CR), the research project RVO:67985815 of the Astronomical Institute AS, and
the Marie Curie PIRSES-GA-2011-295272 RadioSun project. The wavelet analysis
was performed with software based on tools provided by C.~Torrence and
G.~P.~Compo at
\texttt{http://paos.colorado.edu/research/wavelets}.
\end{acknowledgements}

\end{document}